\pdfoutput=1
\documentclass[sigconf]{acmart}
\usepackage{fancyhdr}
\usepackage{amsfonts}
\usepackage{times}
\usepackage{tikz}
\usepackage{tikz}
\usepackage{pgfplots}
\definecolor{bblue}{HTML}{4F81BD}
\definecolor{rred}{HTML}{C0504D}
\definecolor{ggreen}{HTML}{9BBB59}
\definecolor{ppurple}{HTML}{9F4C7C}
\usepackage{comment}
\usepackage{nicefrac}
\usepackage{ifthen}

\usepackage{amsfonts}
\newtheorem{theorem}{Theorem}

\newtheorem{definition}[theorem]{Definition}
\newtheorem{example}[theorem]{Example}

\newtheorem{lemma}[theorem]{Lemma}

\newtheorem{remark}[theorem]{Remark}

\def\bb0{{\mathbb{0}}}

\def\bb{{\mathbf{b}}}

\def\b0{{\mathbf{0}}}

\def\opt{\mathsf{OPT}}
\def\off{\mathsf{OFF}}

\def\b1{{\mathbf{1}}}

\def\bbE{{\mathbb{E}}}

\def\bbN{{\mathbb{N}}}

\def\cA{\mathcal{A}}
\def\cB{\mathcal{B}}

\def\cJ{\mathcal{J}}

\def\cP{\mathcal{P}}

\def\cR{\mathcal{R}}
\def\cS{\mathcal{S}}
\def\cT{\mathcal{T}}

\def\sfs{{\mathsf{s}}}

\def\sf0{{\mathsf{0}}}

\def\nn{\nonumber}

\begin{document}
\title{Scheduling Multi-Server Jobs is Not Easy}
\author{Rahul Vaze}
\email{rahul.vaze@gmail.com}
\affiliation{%
  \institution{Tata Institute of Fundamental Research}
  \city{Mumbai}
  \country{India}
}

\begin{abstract}
The problem of online scheduling of multi-server jobs is considered, where there are a total of $K$ servers, and each job requires concurrent service from multiple servers for it to be processed. Each job on its arrival reveals its processing time, the number of servers from which it needs concurrent service, and an online algorithm has to make scheduling decisions using only causal information with the goal of minimizing the response/flow time. The worst case input model is considered and the performance metric is the competitive ratio. For the case when all job processing time (sizes) are the same, we show that the competitive ratio of any deterministic/randomized algorithm is at least $\Omega(K)$ and propose an online algorithm whose competitive ratio is at most $K+1$. With unequal job sizes, we propose an online algorithm whose competitive ratio is at most $2K \log (K w_{\max})$, where $w_{\max}$ is the maximum size of any job.
With equal job sizes, we also consider the resource augmentation regime where an online algorithm has access to more servers than an optimal offline algorithm. With resource augmentation, we propose a simple online algorithm and show that it has a competitive ratio of $1$ when provided with $2K$ servers with respect to an optimal offline algorithm with $K$ servers. 
\end{abstract}
\maketitle
\section{Introduction} Largely, classical scheduling setups with one or more servers  assume that each  
job requires a single server for processing or a single job can be processed simultaneously by all or a subset of servers using job splitting.
Most of today's data center jobs, however, require service from multiple servers simultaneously for processing
 \cite{app1, app2}. 
This new model is referred to as {\it multi-server job} scheduling that is a paradigm shift from the classical model, where each job {\it blocks} a certain number of servers for its processing, 
and has been an object of immense interest in the recent past \cite{grosof2022wcfs, grosof2022optimal, carastan2019one, brill1984queues, 
zychlinski2020scheduling, zychlinski2023managing, srinivasan2002characterization, grosof2023reset, grosof2023new, harchol2022multiserver, 
wang2021zero, zhao2022learning, hong2022sharp, zhao2023scheduling, olliaro2023saturated, ghanbarian2023performance} 
given its current practical relevance. 


The multi-server job model is typically defined as follows. There are  a total of $K$ servers. Jobs arrive over time and each job on its arrival 
reveals its size (processing time) and the server need (number of servers that it needs for it to be processed at any time). 
The scheduler's job is to select a set of jobs to 
process at any time under the constraint that the sum of the server needs of all the jobs being processed together is at most $K$. A job departs once 
the total service time equal to its size has been dedicated simultaneously from the number of servers it required. The objective is to minimize the sum of the flow time (departure-arrival time) of jobs. The multi server requirement for each job processing brings a new 
combinatorial feature to the scheduling problem, which is generally absent from the classical problems. 

The combinatorial constraint is reminiscent of the well-studied {\it bin packing} problem \cite{bookvaze},  where items with different sizes ($\le 1$) arrive over time and which have to be assigned to bins (with capacity $1$), subject to the constraint that the sum of the size of all assigned items to any bin is at most $1$. The objective is to minimize the number of bins used.  
In fact, using this motivation, a multi-server job model was considered in \cite{shah2008bin}, called {\it bin packing with queues}, where jobs with different sizes arrive over time, and a bin with total size $1$ arrives at each time, and at any time all jobs that can {\it fit} in the bin can be processed together. Jobs not processed in a slot are queued, and the objective is to minimized the expected queue size. Assuming exponential inter-arrival times, and independent and identically distributed job sizes, optimal policy under heavy traffic was established in \cite{shah2008bin}.


In practice, the most widely used scheduling policy for multi-server jobs is FCFS, e.g. CloudSim, iFogSim, EPSim and GridSim cloud
computing simulators \cite{madni2017performance}, or the Google Borg Scheduler \cite{app1}.
even though it is clearly sub-optimal and wastes server capacity, since it can lead to as much as half servers being left idle \cite{you2013comprehensive}. The obvious limitation of FCFS is 
mitigated using the concept of {\it BackFilling} \cite{feitelson1996toward, srinivasan2002characterization, carastan2019one}, where if a job $j$ at the head of the
queue can only be processed at later time $t$ (because of server unavailability till then), then the system allows other jobs that have arrived after job $j$ but 
which can finish before $t$ to preempt job $j$. There are multiple variants of BackFilling; Conservative, EASY, FirstFit. BackFilling avoids the obvious problem with FCFS, however, is still complicated to analyse its flow time performance.

Alternative to BackFilling, another popular algorithm is Most Servers First (MSF) \cite{beloglazov2010energy, maguluri2012stochastic} that preemptively processes the jobs with highest server needs. A more complicated policy is the Max-Weight \cite{maguluri2012stochastic} which searches over all possible packings of jobs in $K$ servers that maximizes (over all packings) the sum of the product of the number of jobs $N_i$ in the system with server requirement of $i$ and the number of jobs with server requirement $i$ that are served by
a packing.	 MSF is a throughput optimal policy, however, its flow time performance is not known. 
Another policy with better throughput performance than MSF is the 
{\it idle-avoid $c-\mu$ rule} \cite{zychlinski2020scheduling, zychlinski2023managing}.

An intuitive policy called {\it ServerFilling} \cite{grosof2022wcfs}, where a set of earliest arrived jobs is selected such that the sum of 
their server requirements is more than $K$. Among this set, jobs are scheduled to be processed in decreasing number of server requirements. When both $K$ and 
all server requirements are a power of $2$, ServerFilling ensures that no server idles. The extension of ServerFilling algorithm when server requirements are not a power of $2$ is called DivisorFilling \cite{grosof2022optimal}.

Most of the performance analysis of the discussed policies for the multi-server job model has been in continuous time where jobs arrive according to a Poisson process with job sizes and server requirements being independent and identically distributed. For this setup, 
ServerFilling is throughput optimal in the heavy traffic limit \cite{grosof2022wcfs}, and it is also shown that the expected flow time of a job with ServerFilling is similar to the expected flow time of a job in a system with a single server implementing FCFS discipline having speed $K$. 
There is also an easy extension of ServerFilling called ServerFilling-SRPT \cite{grosof2022optimal} that has optimal expected flow time performance in the heavy traffic limit. 

When the number of servers required for a job and the system load scales with the total
number of servers, \cite{wang2021zero} considered a scaling regime and obtained results on stability and the probability that an arriving job is blocked, i.e., cannot begin to be process right away has to queue. In this scaling regime, \cite{hong2022sharp} established the first bounds on
mean waiting time in this same asymptotic regime.

A loss model equivalent of multi-server job scheduling has also been studied \cite{arthurs1979sizing, whitt1985blocking, van1989blocking}, where jobs demand a particular service requirement, but if that is not available, are dropped immediately. 

In this paper, in a major departure from prior work, we consider the arbitrary (worst-case) input arrival model, where jobs arrive at arbitrary time instants, with arbitrary 
server requirements and sizes (processing times). 
We consider a discrete time slotted model, where the job arrivals happen at the start of the slot, while departures are accounted at the end of the slots. We consider the online setting, where an algorithm has only causal information about job arrivals and the goal is the flow time minimization, where the decision variables at each slot is the set of jobs to be processed satisfying the constraint that the sum of the server requirement of all processed jobs is at most $K$ (the total number of servers). 

With arbitrary input, the figure of merit for online algorithms is the competitive ratio that is defined as the ratio of the flow time of any online algorithm and the flow time of an optimal offline algorithm that is aware of the full input non-causally and executes an optimal algorithm, maximized over all possible inputs. Thus, an 
online algorithm with small or optimal competitive ratio is robust by definition and has bounded performance for all possible input making it suitable for real-world applications where specific input model is hard to describe. The versatility of this setup is that the optimal offline algorithm need not be known.

When the server needs of all jobs is unity, then the considered problem collapses to the well-studied flow time minimization with $K$ servers, and for which the optimal competitive ratio is $\Theta(\log w_{\max})$ \cite{bookvaze}, where $w_{\max}$ is the maximum size of any job, and is achieved by the multi-server shortest remaining processing time (SRTP) algorithm that at any time processes the $K$ jobs with the shortest remaining time. 

\subsection{Our Contributions}
\begin{itemize}
\item The first result we present is that the competitive ratio of any deterministic/randomized algorithm is at least $\Omega(K)$ for the multi-server jobs problem. This lower bound is derived even when all job sizes are identical. This illustrates the basic  combinatorial difficulty of the considered problem, and essentially a negative result that shows that flow time of any deterministic/randomized algorithm is at least $\Omega(K)$ times the flow time of optimal offline algorithm. We also show that the competitive ratio of ServerFilling is $\Omega(K)$, while that of a greedy algorithm that chooses the largest number of jobs that can be processed together in any slot is arbitrarily large.

\item  For the case when all job sizes are identical, we propose a new algorithm {\bf RA} that schedules the largest set of jobs in increasing order of server requirements in each slot as long as they can occupy all the $K$ servers. Compared to ServerFilling that prefers jobs with larger server requirements, {\bf RA} follows an opposite philosophy of preferring jobs with smaller server requirements, with the motivation of maximizing the number of departures.  We show that the competitive ratio of {\bf RA} is at most $K+1$. 
Intuitively, {\bf RA} appears to keep the difference between the number of remaining jobs with {\bf RA} and the optimal offline algorithm of at most $K$, however, making that intuition concrete in a brute force manner is quite difficult. Hence we present an elegant and simple proof that also exposes some important properties of {\bf RA}.

\item When job sizes are different, we consider the natural generalization of algorithm {\bf RA} and show that its competitive ratio is at most $2K \log (K w_{\max})$, where $w_{\max}$ is the maximum size of any job, while the lower bound on the competitive ratio of any randomized algorithm is $\Omega(\max\{K, \log ( w_{\max})\})$.

\item Given that the competitive ratio of any deterministic or randomized algorithm is at least $\Omega(K)$, we also consider the resource augmentation regime, where 
an online algorithm has access to more resources than the optimal offline algorithm. For this problem, the resource augmentation regime takes the form that an online algorithm has access to more servers than the $K$ servers available for the optimal offline algorithm. For the case when all job sizes are identical, we propose a simple algorithm and show that it has a competitive ratio of $1$ when provided with $2K$ servers with respect to an optimal offline algorithm with $K$ servers. This result has significant system design implications that shows that to get the same performance as the   optimal offline algorithm with $K$ servers, one needs to deploy $2K$ servers in the online paradigm. 
Important question that remains: what is the minimum number of extra servers an online algorithm needs to get the same performance as the optimal offline algorithm with $K$ servers.

\end{itemize}

\section{System Model} We consider a slotted time system with set of jobs $\cJ$ that arrive arbitrarily over time. In particular, job $j \in \cJ$ arrives at slot $a_j$, with size $w_j \in \bbN$ and server requirement of $\sfs_j$. There are a total of $K=2^m$ (for some $m$) servers with unit speed, and job $j$ can be processed during slot $t$ only if $\sfs_j$ servers are assigned to it at slot $t$. For any slot, each server can process at most one job. Following prior work \cite{grosof2022wcfs, grosof2022optimal}, for each job $j\in\cJ$, we let $\sfs_j = 2^a$ for some $a=0, \dots, \log K$, which is well motivated in practice. 
Job $j$ is completed as soon as $\sfs_j$ servers have worked simultaneously for it for $w_j$ slots (possibly over non-contiguous slots). We account for job arrivals at the beginning of a slot, and departures at the end. We consider the model where both preemption (a job's processing can be halted and restarted) and job migration (jobs can be processed by different set of servers in different slots) is allowed. 

In this paper, we consider both cases when all jobs are of same size $w=w_j, \ \forall \ j\in \cJ$, and job sizes are different. For the ease of exposition, we first consider the equal job sizes case, and deal with the unequal job sizes case in Section \ref{sec:unequalsizes}. With equal job sizes, without loss of generality, we let $w=1$ equal to the slot width, and thus a job departs at the end of slot $t$ if it is chosen to be processed in slot $t$. 

An online algorithm $\cA$ at any slot $t$ is aware of all the jobs that have arrived till slot $t$, and makes its scheduling decisions at slot $t$ (deciding the set of jobs $\cP_\cA(t)$ to process at slot $t$ such that $\sum_{j \in \cP_\cA(t)} \sfs_j \le K$) depending on that. Thus, $\cA$ is not aware of $|\cJ|$. 
With $\cA$, let the departure time of job $j$ be $d_j(\cA)$. Then the flow time of job $j$, $f_j(\cA) = d_j(\cA)-a_j$, and the metric of interest is the total flow time 
\begin{equation}\label{eq:probform} 
F_\cA  = \sum_{j\in \cJ} f_j(\cA).
\end{equation}

In comparison, $\opt$ is defined as the optimal offline algorithm that is aware of the full input non-causally and makes optimal scheduling decisions. 
To evaluate the performance of $\cA$, we consider the metric of  competitive ratio that is defined as 
$$\mu_\cA = \max_\sigma \frac{F_\cA(\sigma)}{F_\opt(\sigma)},$$
where $\sigma$ is the input ($(a_j, \sfs_j)_{j=1}^\cJ$). We are not making any assumptions on the input $\sigma$,  which is arbitrary and possibly can be chosen by an adversary,
and the quest is to design online algorithms with small competitive ratios. 
Hereafter, except Section \ref{sec:unequalsizes}, we assume that all job sizes are equal and in particular $w=1$, without repeatedly mentioning it.

\section{WarmUp}\label{sec:warmup} To get a feel for the challenge faced by any online algirithm for solving the considered problem, we begin by lower bounding the competitive ratio of the Server-filling algorithm (SFA) \cite{grosof2022wcfs} that is known to be optimal when the input is stochastic with exponentially distributed inter-arrival times and independently distributed job sizes in the heavy traffic limit.

{\bf SFA:} Let $K$ and $\sfs_j$ be some power of $2$ for all $j$. Then choose the smallest set $\cS$ of earliest arrived jobs such that the sum of $\sum_{j\in \cS}\sfs_j\ge K$. Among this set, choose the jobs to process in decreasing order of $\sfs_j$. Given that $K$ and $\sfs_j$ are some power of $2$ for all $j$, this ensures that no server is idling as long as there is work in the system.

To lower bound the competitive ratio of SFA, we consider the following input.
Let $K/2$ jobs with $\sfs_j=1$ arrive at slot $1$, while one job with $\sfs_j=K$ arrives at  slots $1, \dots, T$. From the definition of SFA, the set $\cS$ chosen by SFA for each slot $1, \dots, T$ contains exactly one job with $\sfs_j=K$ that arrived in that slot itself, making the $K/2$ jobs with $\sfs_j=1$ wait until slot $T$. Hence the flow time of SFA is at least $KT/2$  counting only the flow time of $K/2$ jobs with $\sfs_j=1$ that arrived at  slot $1$.

In comparison, consider an algorithm $\cB$ that processes all $K/2$ jobs with $\sfs_j=1$ in slot $1$, and then processes one job with $\sfs_j=K$ in slots $2, \dots, T+1$. Thus, the flow time of $\cB$ is at most $K/2+2T+1$. Since $\opt$ is as good as $\cB$, the competitive ratio of SFA is at least $\Omega(K)$ choosing $T$ large.  

Let $n_\cA(t)$ be the number of remaining jobs with algorithm $\cA$ at slot $t$. $n_\cA(t)$ is a quantity of interest since $F_\cA = \sum_t n_\cA(t)$.
A non-desirable property which SFA satisfies is that it is possible that 
\begin{equation}\label{eq:SFAGap}
n_{\text{SFA}}(t) - n_{\opt}(t) = \Omega(KT)
\end{equation} for any $T$. This inequality is true for the following input. 
Let $K/2$ jobs with $\sfs_j=1$ arrive at every odd numbered  slot $1, 3, 5, \dots T$ (letting $T$ to be odd), while two jobs with $\sfs_j=K$ arrive at slots $1, 2, 3, 4, \dots, T$. By definition, SFA will process one job with $\sfs_j=K$ in slots $1,2,\dots,T$, making $n_{\text{SFA}}(T) = KT/4$ while it is in fact optimal to 
process the $K/2$ jobs with $\sfs_j=1$ arriving in slot $i$ and $i+2$ together in slot $i+2$ for $i$ odd. This keeps $n_\opt(T) \le T$.

It might appear that the SFA's competitive ratio is large since it prefers to schedule the job with larger $\sfs_j$ and keep $\Omega(K)$ jobs waiting with  $\sfs_j=1$. An alternative is to process as many jobs that can be processed together  in each slot thereby maximizing the departures in each slot. Next, we show that this philosophy performs even worse than SFA.

{\bf Algorithm Greedy:} Process as many jobs as possible in a slot in increasing order of $\sfs_j$.

%
Consider the input where at slot $2\ell+1$, $\ell=0,1, \dots, L_1-1$, one job with $\sfs_j=K$ and two jobs  with $\sfs_j=K/4$ arrive. In addition, at slot $2\ell+2$, $\ell=0,1, \dots, L_1-1$, two jobs with $\sfs_j=K/4$ arrive. 
Greedy by its definition, chooses two jobs with $K/4$ to process until slots $2(L_1-1)+2=2L_1$.
Thus, until slot $2L_1$, with Greedy, all the $L_1$ jobs with $\sfs_j=K$ have not been processed at all. Starting from slot $2L_1$, two jobs of size $\sfs_j=K/2$ arrive at 
$2L_1 + n$, $n=0,1, \dots, L_2-1$. With Greedy all the $L_1$ jobs with $\sfs_j=K$ have to wait until $2L_2$ jobs that arrive after slot $L_1$ are processed. Hence, the flow time of Greedy is at least (only counting the flow time of all the $L_1$ jobs with $\sfs_j=K$) 
$$ L_1 L_2 + L_1(L_1+1)/2.$$

Consider an alternate algorithm $\cB$ that processes the job with $\sfs_j=K$ arriving at slot $2\ell+1$ first and then processes the two jobs with $\sfs_j = K/4$ arriving at slot $2\ell+1$ together with two jobs having  $\sfs_j = K/4$ arriving at slot $2\ell+2$. Thus, with $\cB$, at slot $2L_1$, all jobs that have arrived so far have been processed. Thus, the total flow  of $\cB$ is at most 
$$ 4 L_1 + 2 L_2.$$

Hence, the competitive ratio of Greedy is at least $\min\{L_1, L_2\}$, and hence unboundedly large since $L_1,L_2$ can be chosen arbitrarily. Essentially, Greedy keeps servers idling even when there is outstanding work

One can also show that combining  the features of Greedy and SFA does not result in better competitive ratios.
%

Using the insights that we have developed so far, we next present our first main result of this paper that the competitive ratio of any deterministic algorithm is at least $\Omega(K)$.

\section{Lower Bound}
 
\begin{theorem}\label{thm:lbdet}
 The competitive ratio of any deterministic online algorithm for solving \eqref{eq:probform} is $\Omega(K)$ even if $w_j=1, \ \forall \ j\in \cJ$.
\end{theorem}
\begin{proof}
  Consider any deterministic algorithm $\cA$. For $\cA$, a slot is defined to be {\bf full} if $\cA$ processes a job with $\sfs_j=K$, and a slot is defined to be {\bf wasted}, when $\cA$ processes $K/2$ jobs with $\sfs_j=1$, in that slot.
Consider the following input. Let at slot $1$, one job with $\sfs_j=K$ and $K/2$ jobs with $\sfs_j=1$ arrive. Depending on the action of $\cA$, i.e. choosing a slot as full or wasted, the input is defined as follows.
\begin{enumerate}
\item If slot $t$ is full, then a single job with $\sfs_j=K$ arrives in slot $t+1$. 
\item  If slot $t$ is wasted, then $K/2$ jobs with $\sfs_j=1$ and one job with $\sfs_j=K$ arrives in slot $t+1$.
\end{enumerate}

Let the defined input continue till time $T$. For $\cA$, let $t_1$ be the number of wasted slots chosen by $\cA$ until time $T$. Note that $\cA$ is deterministic, hence the value of $t_1$ is known at slot $1$ itself.

\begin{enumerate}
\item $t_1=\Theta(T)$ or $o(T):$  In this case, consider the following offline algorithm $\off$. $\off$ processes a job in any slot with $\sfs_j=K$ until it has $K$ jobs with $\sfs_j=1$. Whenever there are $K$ jobs with $\sfs_j=1$, $\off$ processes all of them in the same slot. Thus, $\off$ never idles any server, while $\cA$ is wasting half the capacity for $t_1$ slots. Thus, at time $T$, $\cA$ has  $t_1$ remaining jobs with $\sfs_j=K$, while $\off$ has $t_1/2$  remaining jobs with $\sfs_j=K$.  

{\bf Input after time $T$:}  starting from time $ T$, no jobs arrive for time $T+1$ to $T+t_1/2$. 
Thus, at time $T+t_1/2$, $\cA$ has  $t_1/2$ remaining jobs with $\sfs_j=K$, while $\off$ has no remaining jobs.

{\bf Input after time $T+t_1/2$:} Two jobs with $\sfs_j=K/2$ arrive 
at time $T +t_1/2+ \ell$ for $\ell=1, \dots, L$. 

Its best for $\cA$ (in terms of minimizing its flow time) to process the two jobs with 
$\sfs_j=K/2$ in slots $T +t_1/2+1$ to $T+t_1/2+L$ before processing any of the $t_1/2$ outstanding jobs with $\sfs_j=K$ remaining at time $T +t_1/2$. Thus, the flow time of $\cA$  is at least 
(counting only the flow time of $t_1/2$ remaining jobs at time $T$) $\Omega(Lt_1)$.

In comparison, the flow time for jobs processed by $\off$ until time $T$ is $O(KT+ Kt_1) $ and for jobs processed by $\off$ during slot $T+1$ to $T+t_1/2$ is $O(t_1T+t_1^2)$ and after time $T+t_1/2$ is $O(L)$. Thus, the competitive ratio of $\cA$ is $\Omega(t_1)$ by choosing $L=T^2$. Since 
$t_1=\Theta(T)$ or $o(T)$, the competitive ratio of $\cA$ can be made arbitrarily large by choosing $T$ large.

\item $t_1=O(1)$ or $0$. If $t_1=0$ the $K/2$ jobs with $\sfs_j=1$ that arrive in slot $0$ are not processed by $\cA$ till time $T$. Hence, the flow time of $\cA$ for jobs with $\sfs_j=1$ is exactly $K/2\cdot T$. Consider an offline algorithm $\off'$ that processes the $K/2$ jobs in slot $1$ and thereafter processes all the jobs with $\sfs_j=K$ in slots $2$ to $T+1$ has a flow time of at most $T+1+K/2$. Thus, we get that the competitive ratio of $\cA$ is at least $K$. Identical argument works for $t_1=O(1)$.
\end{enumerate}
\end{proof}

We can extend the result of Theorem \ref{thm:lbdet} to randomized online algorithms as well.
\begin{theorem}\label{thm:lbrand}
  The competitive ratio of any randomized online algorithm for solving \eqref{eq:probform} is $\Omega(K)$ even if $w_j=1, \ \forall \ j\in \cJ$.
\end{theorem}
Theorem \ref{thm:lbrand} is proved using Yao's recipe \cite{bookvaze} and provided in the Appendix.

%

{\it Discussion:} The main result obtained in this section, that the competitive ratio of any randomized algorithm is $\Omega(K)$ is effectively a negative result
 and points to the basic difficulty in finding an efficient algorithm to schedule multi-server jobs even when the size of each job is $1$. 
 The main reason for this negative result is the combinatorial aspect of the problem which effectively either makes an algorithm idle some servers even when there is outstanding work or makes it keep a large number of jobs with small server requirements waiting behind jobs with large server requirements. At this point it is not clear whether there exists any algorithm that can achieve the derived lower bound. We saw in Section \ref{sec:warmup} that SFA meets this lower bound but because of SFA satisfying the relation \eqref{eq:SFAGap}, it appears difficult to show that the competitive ratio of SFA is at most $K$ or $O(K)$. 
 The reason for this is that to derive an upper bound on the competitive ratio of any algorithm, we need to upper bound $n_\cA(t) - n_\opt(t)$ at all times $t$. 
 An obvious choice for keeping $n_\cA(t) - n_\opt(t)$ small is to process as many jobs 
 as possible in a single slot while ensuring that all servers are occupied. With this motivation, we next consider a modification of SFA that always occupies the $K$ servers as long as it possible, but prefers jobs with smaller $\sfs_j$'s instead of larger $\sfs_j$'s as with SFA.

\section{Algorithm: {\bf RA}}\label{sec:ra}
Let the size of all jobs be $1$.
At any time $t$, order the remaining jobs in non-decreasing sizes of $\sfs_j$, and in terms of arrival time if their $\sfs_j$'s are the same. Define window sets $S_i(t), i\ge 1$, where $S_i(t)$ 
contains the $i^{th}$ job (in order) and as many consecutively indexed jobs $i+1, i+2, \dots, $ available at time $t$ such that $\sum_{j\in S_i(t)} \sfs_j \le K$, i.e. they fit in one slot for processing.

Algorithm {\bf RA}:  Process all jobs of set $S_{i^\star}(t)$ in slot $t$, where $i^\star = \min \{i :  \sum_{j\in S_i(t)} \sfs_j = K\}$.  $S_{i^\star}(t)$ is the earliest (in order) window set such that all its jobs exactly fit the $K$ servers. If no such set $S_{i^\star}(t)$ exists, process all jobs from set $S_1(t)$ in slot $t$.

The main intuition of algorithm {\bf RA} is that for each slot choose as many jobs to process 
while ensuring that no server idles (as long as it is possible). If there is no subset that can be processed together while making 
all the servers busy, choose that subset that has the largest number of jobs that can be processed together.
We next present a couple of examples to illustrate the working of algorithm {\bf RA}.
\begin{example}\label{exm:RA:1}
  Let $K=8$ and let there be six remaining jobs $\{j_1, \dots, j_6\}$ with $\sfs_j=\{1, 1, 1, 1, 2, 4\}$. Then, by definition, there are 
  six window sets,
  $S_1 = \{j_1, j_2, j_3, j_4, j_5\}$, $S_2=\{j_2, j_3, j_4, j_5\}$, $S_3=\{j_3, j_4, j_5, j_6\}$, $S_4=\{j_4, j_5, j_6\}$, 
  $S_5=\{ j_5, j_6\}$, $S_6=\{j_6\}$.
  Then {\bf RA} processes the {\it last} four jobs $\{j_3, j_4, j_5, j_6\}$ with 
  $\sfs_j's$, $1, 1, 2, 4$ since $S_{i^\star} = S_3$, i.e., earliest indexed window set whose $\sfs_j$'s completely occupy the $K$ servers.
\end{example}
\begin{example}\label{exm:RA:2}
  Let $K=8$ and let there be six remaining jobs $\{j_1, \dots, j_6\}$ with $\sfs_j=\{1, 1, 1, 1, 2, 8\}$. Then, by definition, there are 
  six window sets,
  $S_1 = \{j_1, j_2, j_3, j_4, j_5\}$, $S_2=\{j_2, j_3, j_4, j_5\}$, $S_3=\{j_3, j_4, j_5\}$, $S_4=\{j_4, j_5, \}$, 
  $S_5=\{ j_5\}$, $S_6=\{j_6\}$. 
  Since for $S_1, \dots, S_5$ some servers have to idle, {\bf RA} processes only the last job $\{j_6\}$ with 
  $\sfs_j=8$, since $S_{i^\star} = S_6$, i.e., earliest indexed window set whose $\sfs_j$'s completely occupy the $K$ servers. 
 Note that this action of {\bf RA} appears sub-optimal in terms of minimizing the flow time since there is a possibility of processing five jobs $\{j_1, j_2, j_3, j_4, j_5\}$ together instead of $j_6$. {\bf RA} makes its action towards making sure that no server capacity is wasted while ensuring that largest set of jobs can be processed in each slot.
\end{example}
The main result of this section is as follows.
\begin{theorem}\label{thm:ub}
  The competitive ratio of {\bf RA} for solving \eqref{eq:probform} is at most $K+1$ when $\sfs_j=2^a$ for some $a$, and $w_j=1, \ \forall \ j\in\cJ$.
\end{theorem}
In light of Theorem \ref{thm:lbrand}, {\bf RA} is (order-wise) optimal. Let {\bf RA} be represented as $\cA$ and $n(t)$ be the number of remaining jobs at slot $t$ with $\cA$.
The main intuition behind Theorem \ref{thm:ub} is that it appears that $n(t) \le n_o(t)+ K$,  i.e., number of remaining jobs with $\cA$ are at most $K$ more than 
that of $\opt$ at any point of time. Essentially, $n(t) - n_o(t)$ can become large if large number of jobs with small value of $\sfs_j$ are not processed by $\cA$ but by $\opt$. 
From the definition of the algorithm {\bf RA}, it avoids this situtation since it prefers small number of jobs with large values of $\sfs_j$ to be processed only if the number of jobs with small values of $\sfs_j$ cannot completely fit the $K$ servers. 

To make this intuition concrete, consider a set of slots $[t_1, t_2]$ where for each $t\in [t_1, t_2]$, {\bf RA} processes a single job with $\sfs_j=K$ while $\opt$ processes two jobs $j_{t1}, j_{t2}$ with $\sfs_{j_{t1}},\sfs_{j_{t2}}$. If {\bf RA} also had $j_{t1},j_{t2}$ as its remaining jobs, then the reason that it did not process them together is that $\sfs_{j_{t1}} + \sfs_{j_{t2}} < K$. Thus, the gap $n(t) - n_o(t)$ is growing by $1$ in each slot $t\in [t_1, t_2]$. However, the gap $n(t) - n_o(t)$ cannot increase beyond $K$ since by that time {\bf RA} will get an opportunity to process a set of $j_{t1}, j_{t2}$ jobs in a single slot as soon as 
$\sum_t\sfs_{j_{t1}} + \sfs_{j_{t2}} \ge K$.
One can consider many such similar input instances to verify that $n(t) - n_o(t) \le K$, however,  proving it in a brute force manner runs into combinatorial difficulties. We prove a slightly loose bound of $n(t) - 2n_o(t) \le K-1$,  using a simple and elegant argument as follows. 

\subsection{Proof for Theorem \ref{thm:ub}}
Let the algorithm {\bf RA} be denoted as $\cA$. Job $j$ is defined to belong to class $a$ if $\sfs_j=2^a$ for $a=0, 1, \dots, \log K$. 

Since we account for arrival of new jobs at the start of a slot and departures at the end, we need the following definition. 
  \begin{definition}\label{defn:tminus} For a slot $t$, we let $t^-$ to denote the start of slot, where the set of remaining jobs $R(t^-)$ with any algorithm are i) the remaining jobs from previous slots  and ii) new jobs that arrive in slot $t$. The end of slot $t$ is denoted as $t^+$, where the set of remaining jobs  $R(t^+)$ for any algorithm is $R(t^-)\backslash \cP(t)$, where $\cP(t)$ is  the set of jobs that were processed in slot $t$.
  \end{definition}

For $\cA$, let $R(t^-)$ be the set of outstanding/remaining jobs at slot $t^-$ with $n(t) = |R(t^-)|$, and $n_a(t)$ is the number of remaining jobs among $n(t)$ belonging to class $a$. 
 
   For $\cA$, let $V(t) = \sum_{j \in R(t^-)} \sfs_j$ be the volume (the total outstanding workload) at the start of slot $t$, i.e.,  before processing jobs in slot $t$. 
 Consider the potential function \begin{equation}\Delta V(t) = V(t) - V^\opt(t),\end{equation}  that represents the difference in volume between $\cA$ and the  $\opt$ at start of slot $t$, before processing jobs in slot $t$.


 For any quantity denoted by $Q\in \{V, \Delta V\}$, $Q_{\ge \ell, \le h}$ means the respective quantity when restricted to jobs of classes 
  between $\ell$ and $h$, and $Q_{x} = Q_{\ge x, \le x}$.
  
  \begin{definition} The system is defined to be {\bf full} at slot $t$ if all the $K$ servers are occupied by $\cA$ for processing jobs. The set of slots when the system is full is denoted as $\cT_f$. 
  If the system is not full at slot $t$, then it is defined to be {\bf relaxed}, and the set of slots when the system is relaxed is denoted as $\cT_r$.
  \end{definition}
  
 \begin{lemma}\label{lem:relax}
  If the system is relaxed at slot $t$, i.e. if $t\in \cT_r$, then $n(t)\le K$.
\end{lemma} 
The proof of Lemma \ref{lem:relax} follows from the following simple combinatorial result.
  
  \begin{lemma}\label{lem:packing} Given that for a job $j$, $\sfs_j=2^a$ for some $0\le a \le \log K$, for any set $S$ of jobs with cardinality at least $K$, there exists a subset $S'\subseteq S$ such that $\sum_{j \in S'} \sfs_j = K$.
\end{lemma}
\begin{proof}
 There are at most $\log K+1$ different choices for $\sfs_j$. Thus any set $S$ of jobs with cardinality at least $K$  either has 
 \begin{enumerate}
 \item $2^\ell$ jobs with $\sfs_j=K/2^\ell$ for some $\ell=1, \dots, K$, or 
 \item $K - \sum_{\ell=1}^{K-1} \b1_\ell$ jobs with $\sfs_j=1$,
 \end{enumerate} 
  where $\b1_\ell =1$ if there is a single job with $\sfs_j=K/2^\ell$ in $S$ and $0$ if  there is no job with $\sfs_j=K/2^\ell$ in $S$ for $\ell=1, \dots, K-1$.
 In either case, there exists a subset $S'\subseteq S$ such that $\sum_{j \in S'} \sfs_j = K$.
 \end{proof}

Lemma \ref{lem:relax} now follows since if $n(t)> K$ then the system would be full with $\cA$.


To complement Lemma \ref{lem:relax}, we have the following lemma for bounding the number of outstanding jobs with $\cA$ at slot $t \in \cT_f$.
\begin{lemma}\label{lem:full}
   For $t\in \cT_f$ 
   $$n(t) \le K-1 + 2  n^\opt(t).$$
  \end{lemma}
 
 Next, using Lemma \ref{lem:relax} and \ref{lem:full} and the following simple observations $F_\cA  = \sum n(t) $, and $|\cT_f| + |\cT_r| \le |\cJ|$, where $|\cJ|$ is the total number of jobs with the input and $F_\opt \ge  |\cJ|$ (the total number of jobs), since the size of each job is $1$, we complete the proof of Theorem \ref{thm:ub}. Proof of Lemma \ref{lem:full} is provided thereafter.

\begin{proof}[ Proof of Theorem \ref{thm:ub}]
  \begin{align*}
  F_\cA& \stackrel{(a)}= \sum_t n(t) ,   \\
  &  \stackrel{(b)}= \sum_{t\in \cT_r} n(t)  +  \sum_{t\in \cT_f} n(t) ,   \\
&  \stackrel{(c)}\le \sum_{t\in \cT_r} (K-1) +  \sum_{t\in \cT_f}  (K-1+ 2 n^\opt(t)),   \\
  &  \le (|\cT_f| + |\cT_r|)(K-1) + 2 \sum_{t} n^\opt(t), \\
&  \stackrel{(d)}\le |\cJ| (K-1) + 2 \sum_{t} n^\opt(t) , \\
&  \stackrel{(e)}\le (K-1)F_\opt + 2 \sum_{t} n^\opt(t) , \\
&  =  (K+1)F_\opt
\end{align*}
where $(a)$ follows from the definition of flow time, and $(b)$ follows by partitioning slot into sets $\cT_f$ and $\cT_r$. Lemma \ref{lem:relax} and \ref{lem:full} together imply $(c)$, while $(d)$ follows since  $(\cT_r + \cT_f) \le |\cJ|$, and because trivially $F_\opt \ge |\cJ|$ we get $(e)$.
\end{proof}

Next, we work towards proving Lemma \ref{lem:full}.
\begin{definition}
For some $t\in \cT_f$, let
${\hat t} < t$, be the earliest slot such that $[{\hat t}, t) \in \cT_f$, i.e. for the whole set of slots $[{\hat t}, t)$ all servers are occupied with $\cA$.
  During interval $[{\hat t}, t)$, the latest slot at which a job belonging to class greater than $a$ is processed is defined as $t_a$. 
  We let $t_a = {\hat t}-1$, if no job with class greater than $a$ is processed in $[{\hat t}, t)$.
  \end{definition}
  
  With these definitions, we have the following intermediate result.
  \begin{lemma}\label{chap:mssched:lem:deltav1}
  For $t\in \cT_f$, $\Delta V_{\le a}(t) \le  \Delta V_{\le a}(t_a+1)$.
\end{lemma}
This result means is that the difference in the volume between $\cA$ and $\opt$ for jobs with class at most $a$ does not increase from the start of slot $t_a+1$ to 
the start of slot  $t$. 
\begin{proof}
Since $t_a+1\ge {\hat t}$, $[t_a+1, t) \in \cT_f$, i.e., all servers are occupied throughout the interval $[t_a+1, t)$ with $\cA$. Moreover, $\cA$ reduces the volume $V_{\le a}$ by maximal 
amount of $K$ at any slot in  $[t_a+1, t)$, since for $t\in\cT_f$, the set of processed jobs $\cP(t)$ in slot $t$ satisfies $\sum_{j\in \cP(t)} \sfs_j=K$.
Hence the reduction in $V_{\le a}$ because of $\cA$ in interval $[t_a+1, t)$
is $K(t-t_a-2)$. 

$\opt$ on the other hand need not have all servers occupied during $[t_a, t)$ or may be work on jobs with classes more than $a$. Hence the total reduction it can achieve for $V^\opt_{\le a}$ in interval $[t_a+1, t)$ is upper bounded by 
$K(t-t_a-2)$. Thus, $\Delta V_{\le a}(t) \le \Delta V_{\le a}(t_a+1)$.
\end{proof}

\begin{lemma}\label{chap:mssched:lem:deltav2}
For $t\in \cT_f$, $\Delta V_{\le a}(t_a+1) \le K-1$.
\end{lemma}
This result implies that the difference of volume between $\cA$ and $\opt$ at the start of slot $t_a+1$ is at most $K-1$.
\begin{proof}

Case I: $t_a={\hat t}-1$. Thus, no job with class more than $a$ is processed by $\cA$ in $[{\hat t}, t)$. Since ${\hat t}-1\in \cT_r$, we get that the total volume of jobs with $\cA$ at both the start and end of slot 
${\hat t}-1$ with class at most $a$ is at most $K-1$. Thus, $V_{\le a}(t_a^+) = \sum_{j \in R(t^+)} \sfs_j \le K-1$.
Moreover, the set of newly  arriving jobs in slot $t_a+1$ is identical for both $\cA$ and the $\opt$. Thus, $\Delta V_{\le a}(t_a+1) \le V_{\le a}(t_a^+)\le K-1$.


Case II: $t_a>{\hat t}-1$. If $\cA$ is processing a job of class more than $a$ at slot $t_a$ this means that the total volume of jobs at slot $t_a^-$ with class at most $a$ is at most $K-1$.\footnote{It is in fact at most $K/2^{a}-1$ but given that we are going to consider all classes, it is sufficient to consider the weakest bound.} This is true since otherwise $\cA$ would have processed a subset of jobs with class at most $a$ at slot $t_a\in \cT_f$ while fully occupying the $K$ servers as it prioritises jobs with smaller $\sfs_j$'s as long as all $K$ servers can be occupied.  Therefore, we get that 
  $V_{\le a}(t_a^+)\le K-1.$
  Moreover, the set of newly  arriving jobs in slot $t_a+1$ is identical for both $\cA$ and the $\opt$, thus, we get $\Delta V_{\le a}(t_a+1) \le V_{\le a}(t_a^+)\le K-1$.

\end{proof}

Combining Lemma \ref{chap:mssched:lem:deltav1} and \ref{chap:mssched:lem:deltav2}, we get the following result.

\begin{lemma}\label{lem:driftequalsize}
  For $t\in \cT_f$, $\Delta V_{\le a}(t) \le  K-1$.
\end{lemma}

Using Lemma \ref{lem:driftequalsize}, we complete the proof of Lemma \ref{lem:full}, that connects the number of remaining jobs with $\cA$ and the $\opt$, belonging to a certain set of classes.

\begin{proof}[Proof of Lemma \ref{lem:full}]
\begin{align}\nn
   \sum_{a=a_1}^{a_2} n_a(t)& \stackrel{(a)}= \sum_{a=a_1}^{a_2}\frac{V_a(t)}{2^a},  \\\nn
  &\stackrel{(b)} =  \sum_{a=a_1}^{a_2}\frac{\Delta V_a(t) +  V^\opt_a(t)}{2^a}, 
    \end{align}
  \begin{align}\nn
  &=  \sum_{a=a_1}^{a_2}\frac{\Delta V_{\le a}(t)  - \Delta V_{\le a-1}(t)}{2^a} + \sum_{a=a_1}^{a_2}\frac{ V^\opt_a(t)}{2^a},  \\\nn
    & \stackrel{(c)}\le    \frac{\Delta V_{\le a_2}(t)}{2^{a_2}} + \sum_{a=a_1}^{a_2-1}\left(\frac{\Delta V_{\le a}(t) }{2^{a}}- \frac{\Delta V_{\le a}(t) }{2^{a+1}}\right) - 
  \frac{\Delta V_{\le a_1-1}(t)}{2^{a_1}} \\ \nn
 & \quad \quad  + \sum_{a=a_1}^{a_2}\frac{ V^\opt_a(t)}{2^a}, \\ \nn
  &  \stackrel{(d)}\le    \frac{\Delta V_{\le a_2}(t)}{2^{a_2}} + \sum_{a=a_1}^{a_2-1}\frac{\Delta V_{\le a}(t) }{2^{a+1}} - 
  \frac{\Delta V_{\le a_1-1}(t)}{2^{a_1}} + 2 n^\opt_{\ge a_1, \le a_2}(t), \\ \nn
     &  \stackrel{(e)}\le (K-1)\frac{1}{2^{a_2}} + (K-1) \sum_{a=a_1}^{a_2-1} \frac{1}{2^{a+1}} + 
  \frac{ V^\opt_{\le a_1-1}(t)}{2^{a_1}} + 2 n^\opt_{\ge a_1, \le a_2}(t), \\\nn
   &  \stackrel{(f)}\le c_{a_1,a_2}(K-1)  + 2n^\opt_{ \le a_1-1}(t) + 2 n^\opt_{\ge a_1, \le a_2}(t), \\\nn
   &  \le (K-1) + 2 n^\opt_{\le a_2}(t)
\end{align}
where $(a)$ follows from the definition of $V_a(t)$ as the total remaining volume of jobs  belonging to class $a$ at slot $t$, and $\sfs_j=2^a$ for job $j$ of class $a$, while $(b)$ follows from the definition of $\Delta V_a = V_a - V^\opt_a$. To get $(c)$ we separate the telescopic sum over $a_1$ to $a_2$ into three parts, 
$a_2$, $a_1$ to $a_2-1$ and $a_1-1$. On the second part we use the fact that $\Delta V_{\le a}(t)\le 2 \Delta V_{\le a-1}(t)$. Inequality $(d)$ follows from the definition of $V^\opt_a$, and the fact that the volume of each job in class $a$ is $2^a$. Inequality $(e)$ follows by applying Lemma \ref{lem:driftequalsize} on the first two terms separately, and for the third term use the property that  $- \Delta V_{\le a_1-1}(t) \le  V^\opt_{\le a_1-1}(t)$. Letting $c_{a_1,a_2} = \frac{1}{2^{a_2}} + \sum_{a=a_1}^{a_2-1} \frac{1}{2^{a+1}} \le 1$, and using the fact that volume of a job of class $a$ is $2^a$  on the third term of $(e)$, we get inequality $(f)$.

Letting $a_1=0$ and $a_2=\log K$, and noting that $n(t) =   \sum_{a=0}^{\log K} n_a(t)$, we get the result.
\end{proof}
\vspace{-0.1in}
{\it Discussion:}
In this section, we proposed a simple algorithm {\bf RA} that prefers jobs with smaller $\sfs_j$'s as long as they can completely occupy the $K$ servers and showed that its competitive ratio is at most $K+1$. {\bf RA}'s philosophy is in contrast to the well-known ServerFilling (SFA) algorithm that prefers jobs with larger $\sfs_j$'s while not wasting any server capacity. In Section \ref{sec:warmup}, we showed that the competitive ratio of SFA is at least $\Omega(K)$ similar to the lower bound (Theorem \ref{thm:lbdet}) that we obtain for all deterministic algorithms. It is possible that SFA also has a competitive ratio of at most $K$, however, in light of relation \eqref{eq:SFAGap} that SFA satisfies, it appears difficult to show that. The advantage of {\bf RA} is that it maximizes 
the number of departures (while not wasting server capacity) that directly reflects in keeping the number of remaining jobs low, whose sum is the flow time. 
Because of this, the proof of Theorem \ref{thm:ub} is elegant and exposes the structural properties of {\bf RA} in Lemma \ref{lem:relax} and \ref{lem:full}.

Combining Theorem \ref{thm:lbrand} and \ref{thm:ub}, we conclude that the considered problem is challenging for any online algorithm, and the competitive ratio is $\Theta(K)$, i.e., increases 
linearly in the number of total servers. Under such limitation, the usual extension in the online algorithms literature is to consider the {\it resource augmentation} regime \cite{phillips1997optimal,anand2012resource,bookvaze}, where an online algorithm is allowed use of more resources than the optimal offline algorithm. The hope is that the serious limitation of any 
online algorithm can be overcome with more resources. We consider this in the next section, and show that a simple algorithm with $2K$ servers has the same flow time performance as the offline optimal algorithm with $K$ servers.


\section{Resource Augmentation}
In this section, we consider the resource augmentation regime,  important from a system design point of view, that
explores the possibility of whether online algorithms with constant competitive ratios are possible when given extra resources compared to the $\opt$. We consider an enhancement of algorithm {\bf RA}, called {\bf RA-E}, and show that it has a competitive ratio of $1$ when it is allowed to use $2K$ servers in comparison to $\opt$ that is only allowed $K$ servers.
Essentially, what this means is that to get the same performance as $\opt$ with $K$ servers, algorithm {\bf RA-E} needs $2K$ servers. This is remarkable result, since we 
do not know what $\opt$ is even for $K$ servers.
\vspace{-0.1in}
\subsection{Algorithm {\bf RA-E}}
Let the number of servers be $2K$. The first set of $K$ servers is called {\bf reserved} set, while the second set of $K$ servers is defined to be the {\bf free} set. 

Recall the definition of window sets from algorithm {\bf RA}: Let $R(t)$ be the set of remaining jobs at  slot $t$. Order the jobs in $R(t)$ in non-decreasing sizes of 
$\sfs_j$ and in terms of arrival time if their $\sfs_j$'s are the same. Define window sets $S_i(t), i\ge 1$, where $S_i(t)$ contains the $i^{th}$ job (in order) and as many consecutively indexed jobs $i+1, i+2, \dots, $ available at slot $t$ such that $\sum_{j\in S_i(t)} \sfs_j \le K$, i.e. they fit in one slot for processing. 

Let $i^\star = \min \{i :  \sum_{j\in S_i(t)} \sfs_j = K\}$.  $S_{i^\star}(t)$ is the earliest (in order) window set such that all its job exactly fit the $K$ servers. 

If $i^\star$ exists, then define ${\bar S}(t) = R(t) \backslash S_{i^\star}(t)$, otherwise ${\bar S}(t) = R(t) \backslash S_{1}(t)$. 
Similar to window sets of $S_i(t)$, define the ordered window sets  
${\bar S}_i(t)$ for ${\bar S}(t)$, and define ${\bar i}^\star = \min \{i :  \sum_{j\in {\bar S}_i(t)} \sfs_j \le K\}$. If no such ${\bar i}^\star$ exists, then $S_{{\bar i}^\star}(t) = {\bar S}_1(t)$. Note that compared to  $i^\star$, with ${\bar i}^\star$, because of the inequality in its definition, the full $K$ servers needs not be occupied by jobs in $S_{{\bar i}^\star}$. 


Algorithm {\bf RA-E} does the following:
\begin{enumerate}
\item If $i^\star$ exists, then process jobs from set $S_{i^\star}(t)$ on the reserved set of servers, and  process jobs from set $S_{{\bar i}^\star}(t)$ on the set of free servers.
\item If $i^\star$ does not exist, then process jobs from set $S_{1}(t)$ on the set of reserved servers, and process the smallest job $j_{\min}$ in terms of $\sfs_j$ of $R(t)$ that is not part of $S_{1}(t)$ on the set of free servers. 
\end{enumerate}
Note that unlike the previous sections, in this section, we do not need to enforce that $\sfs_j=2^a$ for some $a$, and $\sfs_j \in [1:K]$.
\begin{remark} In light of Lemma \ref{lem:packing}, the second condition of {\bf RA-E} is effective only if $\sfs_j \ne 2^a$ for some $a$, since otherwise, there is no such job $j_{\min}$.
\end{remark}
 \begin{example}\label{exm:RAE:1}
  Similar to Example \ref{exm:RA:1}, let $K=8$ and let there be six remaining jobs $\{j_1, \dots, j_6\}$ with $\sfs_j=\{1, 1, 1, 1, 2, 4\}$, with six window sets,
  $S_1 = \{j_1, j_2, j_3, j_4, j_5\}$, $S_2=\{j_2, j_3, j_4, j_5\}$, $S_3=\{j_3, j_4, j_5, j_6\}$, $S_4=\{j_4, j_5, j_6\}$, 
  $S_5=\{ j_5, j_6\}$, $S_6=\{j_6\}$. $S_{i^\star} = S_3$. Removing the set $S_3$ from remaining jobs, we have ${\bar S}(t) = \{j_1, j_2\}$. Then clearly, ${\bar S}_{i^\star} = \{j_1, j_2\}$. Thus, {\bf RA-E} processes set $S_3$ on the $K=8$ reserved servers, and $\{j_1, j_2\}$ on the $2$ of the $K=8$ free servers.
\end{example}
  \begin{example}\label{exm:RAE:2}
  In this example, we illustrate the case when the second condition of  {\bf RA-E} is effective on account of  $\sfs_j \ne 2^a$ for some $a$.
  Let $K=8$ and let there be five remaining jobs $\{j_1, \dots, j_5\}$ with $\sfs_j=\{1, 1, 1, 3,  6\}$, with six window sets,
  $S_1 = \{j_1, j_2, j_3, j_4\}$, $S_2=\{j_2, j_3, j_4\}$, $S_3=\{j_3, j_4\}$, $S_4=\{j_4,\}$, 
  $S_5=\{j_5\}$. Since no $i^\star$ exists, $S_1=\{j_1, j_2, j_3,j_4\}$ is processed on the $6$ out of  $K=8$ reserved servers, and $\{j_5\}$ is processed on the $6$ out of $K=8$ free servers, ensuring that at least $K$ servers are busy as long as there is a subset of jobs that can occupy $K$ servers.
 \end{example}
 
We have the following important result for algorithm {\bf RA-E}.
\begin{theorem}\label{thm:aug}
  The flow time of algorithm {\bf RA-E} with $2K$ servers is at most the flow time of $\opt$ with $K$ servers when $\sfs_j \in [1:K]$ and $w_j = 1 \ \forall \ j \in \cJ$.
\end{theorem}
Towards proving this result, we need the following definition.
\begin{definition}\label{}
  Let the number of jobs completely processed (departed) by an online algorithm $\cA$  with input sequence $\sigma_1$ by slot $t$ be $r_t(\sigma_1)$. 
  For any augmented input $\sigma_2$ such that $\sigma_1 \subseteq \sigma_2$, if $\cA$ satisfies the condition that $r_t(\sigma_2) \ge r_t(\sigma_1)$ for all slots $t$, then $\cA$ is defined to satisfy {\bf augmentation} property.
\end{definition}

\begin{lemma}\label{lem:aug}
  Algorithm {\bf RA-E} satisfies the augmentation property.
\end{lemma}
\begin{proof}
  To prove this Lemma, we consider two inputs $\sigma_1$ and $\sigma_2$, where with $\sigma_2$  one additional job $k$ with $\sfs_k$ arrives at slot $a_k$. For the ease of exposition, we suffix the input $\sigma_1$ or $\sigma_2$ to the respective sets of interest.
  We will show that the number of departures by any slot with {\bf RA-E} when input is $\sigma_1$ is at least as many as when input is $\sigma_2$ for $t\ge a_k$.
  Consider slot $t=a_k$. 
  
 Case I: With $\sigma_1$, let $i^\star$ exists at slot $t$. Then with $\sigma_1$, the two-tuple $(S_{i^\star,\sigma_1}(t), S_{{\bar i}^\star,\sigma_1}(t))$ is the set of jobs processed on the reserved and the free set of servers. 
  In this case, with $\sigma_2$, the newly arrived job $k$ can disturb set $S_{i^\star,\sigma_1}(t)$ only if its $\sfs_k$ is smaller than $\sfs$ of some job that is part of $S_{i^\star,\sigma_1}(t)$. Moreover, since $S_{i^\star,\sigma_1}(t)$ exists when input is $\sigma_1$, we also get that $S_{i^\star,\sigma_2}(t)$ exists even when input is $\sigma_2$. Combining these two facts together, 
  the number of jobs processed by {\bf RA-E} $|S_{i^\star,\sigma_2}(t)|$ over the reserved set of servers is such that $|S_{i^\star,\sigma_2}(t)| \ge  |S_{i^\star,\sigma_1}(t)|$.
 
 Next, we consider how $S_{{\bar i}^\star}(t)$ can change. If $S_{i^\star,\sigma_2}(t)=S_{i^\star,\sigma_1}(t)$, then job $k$ can only make 
 $|S_{{\bar i}^\star,\sigma_2}(t)| \ge |S_{{\bar i}^\star,\sigma_1}(t)|$. 
 
 Otherwise, any job $j \in S_{i^\star,\sigma_1}(t)$ but $j \notin S_{i^\star,\sigma_2}(t)$ can either become part of $S_{{\bar i}^\star,\sigma_2}(t)$ by ejecting at most one element of 
 $S_{{\bar i}^\star,\sigma_1}(t)$ or not be part of $S_{{\bar i}^\star,\sigma_2}(t)$. In both cases, $|S_{{\bar i}^\star,\sigma_2}(t)| \ge |S_{{\bar i}^\star,\sigma_1}(t)|$. Thus, in both cases the augmentation property is satisfied.
  
 Case II: With $\sigma_1$, let $i^\star$ does not exist at slot $t$. Then with $\sigma_1$, $(S_1, \{j_{\min}\})$ is the set of jobs processed on reserved and free set of servers. In this case, with $\sigma_2$, job $k$ can disturb $S_1$ by becoming a part of 
  $S_1$ either by ejecting an existing element or without ejecting any existing element. If an element $e$ is ejected, then $j_{\min}$ can either remain as it is or become $e$.
  In either case, the number of jobs departing at slot $t$ does not 
  decrease with $\sigma_2$ compared to $\sigma_1$.
  
 The same holds for any slot $t\ge a_k$, and by incrementally adding one job at a slot that is part of $\sigma_2$ 
 but not of $\sigma_1$, we get the result.
\end{proof}

\begin{lemma}\label{lem:work}
  Algorithm {\bf RA-E} with $2K$ servers does as much work as $\opt$ with $K$ servers by any slot $t$.
\end{lemma}
\begin{proof}
Consider any slot $t$ and let $i^\star$ exist at slot $t$. Then the two-tuple $(S_{i^\star}(t), S_{{\bar i}^\star}(t))$ is the set of jobs processed on reserved and free set of servers, respectively, and by definition of $S_{i^\star}(t)$:  $\sum_{j\in S_{i^\star}(t)} \sfs_j=K$. Thus, at least $K$ servers are busy at slot $t$.

If $i^\star$ does not exist at slot $t$, $(S_1, \{j_{\min}\})$ is the set of jobs processed on reserved and free set of servers. In this case, $\sum_{j\in S_1} \sfs_j$ can be less than $K$, but by definition $\sum_{j\in S_1} \sfs_j + \sfs_{j_{\min}} \ge K$ since otherwise $j_{\min}$ would have been part of $S_1$. Thus, at least $K$ servers are busy at slot $t$.

The only way at least $K$ servers are not busy with {\bf RA-E} is when $\sum_{j\in R(t)} \sfs_j < K$, where $R(t)$ is the set of remaining jobs. But in this case, 
{\bf RA-E} finishes all the work in slot $t$. 

In comparison, $\opt$ has access to at most $K$ servers, and hence can only keep them active at any slot. Thus, {\bf RA-E} with $2K$ servers does as much work as $\opt$ with $K$ servers by any slot $t$.
\end{proof}

\begin{proof}[Proof of Theorem \ref{thm:aug}] Let the full input job sequence be $\sigma$. Let the subset of input $\sigma_t\subseteq\sigma$ be the set of jobs that the $\opt$ with $K$ servers finishes completely by slot $t$, when the input sequence is $\sigma$. 
Therefore from Lemma \ref{lem:work}, if the input sequence is just $\sigma_t$,
{\bf RA-E}  finishes all $|\sigma_t|$ jobs by slot $t$ with $2K$ servers. 

Now we make use of the augmentation property. Let $\sigma_t = \sigma_1$ and $\sigma = \sigma_2$, then 
the augmentation property of {\bf RA-E} (Lemma \ref{lem:aug}) implies that for any slot $t$, at least $|\sigma_t|$ jobs will be completed by slot $t$ with {\bf RA-E} even when the input sequence is $\sigma$.

This implies that for any $k$, the departure time of the $k^{th}$ job with {\bf RA-E} is no later than  the departure time of the  $k^{th}$ job with the $\opt$ for any job arrival sequence $\sigma$. Note that the order of departure of jobs with the {\bf RA-E} and $\opt$ might be different. Thus, we get that 
\begin{equation}\label{eq:finaugbound}
\sum_{j \in \cJ} d_j(\opt) \ge   \sum_{j \in \cJ} d_j({\bf RA-E}).
\end{equation}

Recall that the flow time is $\sum_{j \in \cJ} (d_j - a_j)$ and $\sum_{j \in \cJ} a_j$ is independent of the algorithm. Thus, to claim that $F_{\opt} \ge F_{\cA}$, it is sufficient to show that  $\sum_{j \in \cJ} d_j(\opt) \ge  \sum_{j \in \cJ} d_j(\cA)$ as done in \eqref{eq:finaugbound}. Hence the proof is complete. 
\end{proof}
{\it Discussion:} In this section, we showed that a simple algorithm {\bf RA-E}, that enjoys the augmentation property and ensures that $K$ servers are busy as long as there is sufficient work in the system has the same flow time performance with $2K$ servers as that of the optimal offline algorithm with $K$ servers. Given that without resource augmentation, the competitive ratio of any randomized algorithm is $\Omega(K)$, this is a remarkable result, and 
recovers all the lost power because of online-ness of the algorithms compared to offline optimal algorithm. One question that remains to be answered: 
what is the least number of extra servers needed to get $r$-competitive online algorithm, or what is the best efficiency ratio (ratio of number of servers available with an online algorithm and the $\opt$) to obtain a $r$-competitive online algorithm. What we have shown is that {\bf RA-E} is $1$-competitive with an efficiency ratio of $2$.

\section{Jobs with unequal sizes}\label{sec:unequalsizes}
So far in this paper we have only dealt with the case when all job sizes are equal, i.e., $w_j=w_k$ for any $j\ne k$.  In this section, we consider the general case when job sizes are unequal. In particular, we assume that size of job $j$ is $w_j \in \bbN$ to fit the slotted time model, and job $j$ departs as soon as $\sfs_j$ servers have processed it simultaneously for $w_j$ slots (possibly over non-contiguous slots). For this model, we next propose an extension of the {\bf RA} algorithm and bound its competitive ratio. Similar to Section \ref{sec:ra}, we require that the server need of each job $\sfs_j=2^a$ for some $a$.

{\bf RA-Size}: Let the effective size of job $j$ be $w'_j =  w_j \cdot \sfs_j$. The remaining effective size of job $j$ at time $t$ is $w'_j(t) = w_j(t) \cdot \sfs_j$, where 
$w_j(t)$ is the remaining size of job $j$ at time $t$. Thus, $w'_j =  w_j \cdot \sfs_j = w_j(a_j) \cdot \sfs_j $.

At at time slot $t$, order the remaining jobs in increasing order of the remaining effective sizes $w'_j(t)$ of the jobs and in terms of arrival time if their $\sfs_j$'s are the same.
Define window sets $S_i(t), i\ge 1$, where $S_i$ 
contains the $i^{th}$ job (in order) and as many consecutively indexed jobs $i+1, i+2, \dots, $ available at time $t$ such that $\sum_{j\in S_i(t)} \sfs_j \le K$, i.e. 
they fit in one slot for processing. Rest of the algorithm is the same as {\bf RA}. Let $1\le w_j \le w_{\max}$, then we have the following result on the competitive ratio of {\bf RA-Size}

\begin{theorem}\label{thm:ubunequalsize}
  The competitive ratio of {\bf RA-Size} is at most $(K+1)\log (Kw_{\max})$ when $\sfs_j=2^a$ for some $a$.
\end{theorem}
\begin{remark}
When job sizes are equal, algorithm {\bf RA-Size} is same as {\bf RA}, however, it is worth noting that Theorem \ref{thm:ub} is not a special case of Theorem \ref{thm:ubunequalsize}, since with equal job sizes $w_{\max}=1$, and Theorem \ref{thm:ubunequalsize} implies a competitive ratio of $(K+1)\log (K)$, while Theorem \ref{thm:ub}'s competitive ratio bound of $K+1$ for {\bf RA} is significantly better. This 
difference is a result of analytical difficulty in proving Theorem \ref{thm:ubunequalsize} with unequal job sizes.
\end{remark}

Recall that when all $\sfs_j=1$, the considered problem \eqref{eq:probform}  is the classical flow time minimization problem with $K$ servers, for which $\Omega(\log (w_{\max}))$ is a lower bound on the competitive ratio of any randomized algorithm \cite{bookvaze}. Combining the lower bound that we have derived in Theorem \ref{thm:lbrand}, we get that for problem \eqref{eq:probform}, $\Omega(\max\{K, \log (w_{\max})\})$ is a  lower bound on the competitive ratio of any randomized algorithm. Compared to this lower bound, 
the upper bound on the competitive ratio of {\bf RA-Size} derived in Theorem \ref{thm:ubunequalsize} is off by a multiplicative factor. At this point it is not obvious, whether the lower or the upper bound is loose and this resolution is left for future work.

\begin{remark}
When job sizes are different, {\bf RA-Size}'s extension with resource augmentation (similar to {\bf RA}'s extension to {\bf RA-E}) does not yield a result like Theorem \ref{thm:aug}.
\end{remark}

\section{Numerical results}\label{sec:sim}
In this section, we present simulation results for the flow time (per job). 
Presenting simulation results in the worst case input model is challenging, since the $\opt$ is unknown. Thus for the purposes of 
comparison, we use the ServerFilling algorithm that has been shown to have near-optimal performance with stochastic inputs.
For all simulations, with arrival rate 
$\textsf{arr}$, we generate $\textsf{arr}$ jobs on average per slot and then distribute them over the time horizon arbitrarily. Also for simplicity, we let 
all job sizes to be equal.
We let the jobs to arrive for $100$ slots, and then let the time horizon to be earliest time by which all jobs are complete, and 
compute the per-job flow time by averaging over large number of realizations. 

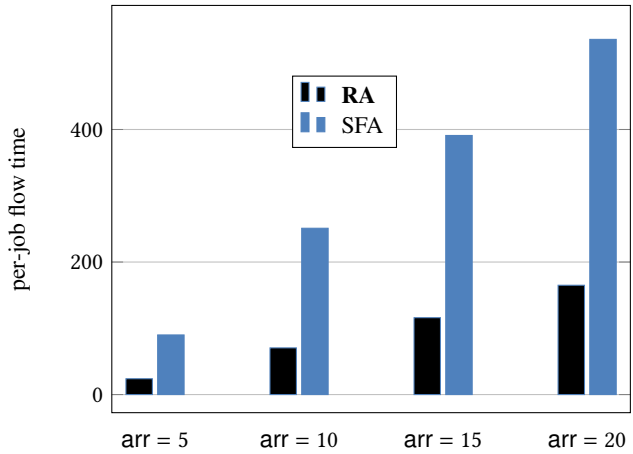
\begin{figure}
\centering
\begin{tikzpicture}
    \begin{axis}[
        width  = \columnwidth,
        height = 7cm,
        major x tick style = transparent,
        ybar,
        bar width=10pt,
        ymajorgrids = true,
        ylabel = {per-job flow time},
        symbolic x coords={$\textsf{arr}=5$, $\textsf{arr}=10$, $\textsf{arr}=15$, $\textsf{arr}=20$},
        xtick = data,
        scaled y ticks = false,
        legend cell align=left,
        legend style={
                at={(.55,.65)},
                anchor=south east,
                column sep=1ex}
    ]

     \addplot[style={bblue,fill=black,mark=none}]
            coordinates {($\textsf{arr}=5$, 23.9) ($\textsf{arr}=10$,70.45) ($\textsf{arr}=15$,116) ($\textsf{arr}=20$, 165)};

        \addplot[style={bblue,fill=bblue,mark=none}]
            coordinates {($\textsf{arr}=5$, 90) ($\textsf{arr}=10$, 251) ($\textsf{arr}=15$,391) ($\textsf{arr}=20$,536)};

        \legend{{\bf RA},  SFA}
    \end{axis}
\end{tikzpicture}
\caption{Comparison of per-job flow time for $K=16$ and $\sfs$ uniformly randomly among $[1 \ 2\ 4 \ 8 \ 16]$.}
\label{fig:K16} 
\end{figure}

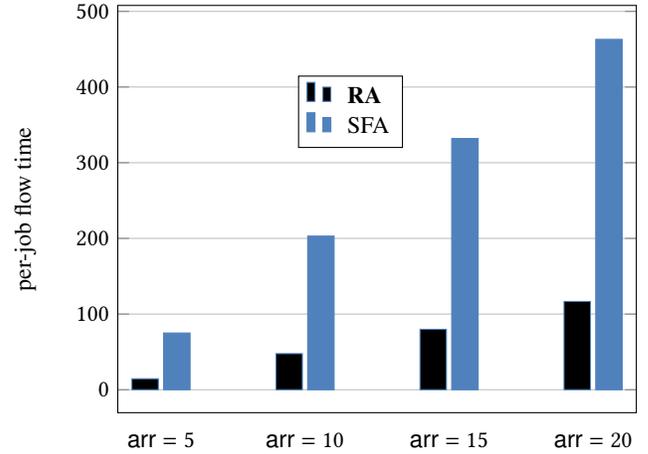
\begin{figure}
\centering
\begin{tikzpicture}
    \begin{axis}[
        width  = \columnwidth,
        height = 7cm,
        major x tick style = transparent,
        ybar,
        bar width=10pt,
        ymajorgrids = true,
        ylabel = {per-job flow time},
        symbolic x coords={$\textsf{arr}=5$, $\textsf{arr}=10$, $\textsf{arr}=15$, $\textsf{arr}=20$},
        xtick = data,
        scaled y ticks = false,
        legend cell align=left,
        legend style={
                at={(.55,.65)},
                anchor=south east,
                column sep=1ex}
    ]

     \addplot[style={bblue,fill=black,mark=none}]
            coordinates {($\textsf{arr}=5$, 14.5) ($\textsf{arr}=10$, 47.71) ($\textsf{arr}=15$, 79.89) ($\textsf{arr}=20$, 116.670)};

        \addplot[style={bblue,fill=bblue,mark=none}]
            coordinates {($\textsf{arr}=5$, 75) ($\textsf{arr}=10$,203) ($\textsf{arr}=15$,332) ($\textsf{arr}=20$,463)};

        \legend{{\bf RA},  SFA}
    \end{axis}
\end{tikzpicture}
\caption{Comparison of per-job flow time for $K=32$ and $\sfs$ uniformly randomly among $[1 \ 2\ 4 \ 8 \ 16 \ 32]$.}
\label{fig:K32} 
\end{figure}

With total $K$ servers, we start with the simplest setting, where each job has its server need $\sfs$ uniformly distributed among
$\{2^\ell : \ell=0, \dots, \log K\}$  and plot the 
per-job flow time in Figs. \ref{fig:K16} and \ref{fig:K32} for $K=16$ and $K=32$, where we compare the performance of {\bf RA} and SFA as a function of arrival rate $\textsf{arr}$. 
From Figs. \ref{fig:K16} and \ref{fig:K32}, we observe that {\bf RA} has significantly better performance than SFA for all arrival rates. Recall that SFA has close to optimal performance in the stochastic case where the load (average server needs of all jobs arriving in a single slot is less than $K$), which is also necessary for ensuring {\it stability}. In the worst-case input, load need not be less 
than $K$, and as we can see,  {\bf RA} has significantly  better performance than SFA for larger loads.

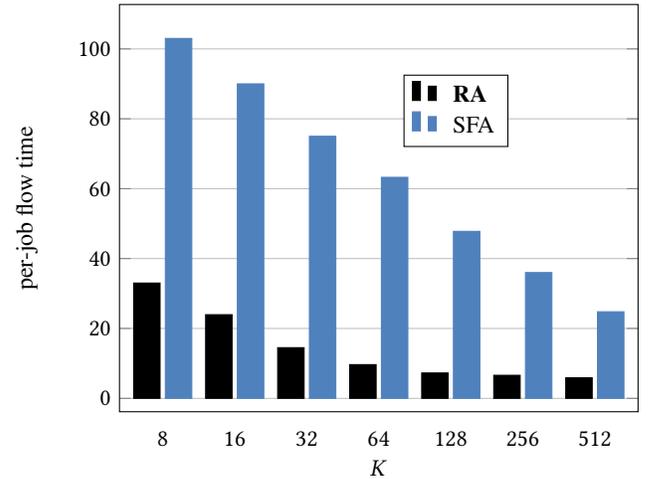
\begin{figure}
\centering
\begin{tikzpicture}
    \begin{axis}[
        width  = \columnwidth,
        height = 7cm,
        major x tick style = transparent,
        ybar,
        bar width=10pt,
        ymajorgrids = true,
        ylabel = {per-job flow time},
        xlabel = {$K$},
        symbolic x coords={$8$, $16$, $32$, $64$, $128$, $256$, $512$},
        xtick = data,
        scaled y ticks = false,
        legend cell align=left,
        legend style={
                at={(.75,.65)},
                anchor=south east,
                column sep=1ex}
    ]

     \addplot[style={black,fill=black,mark=none}]
            coordinates {($8$, 33) ($16$, 23.94) ($32$, 14.5) ($64$, 9.65) ($128$, 7.26) ($256$, 6.6) ($512$, 5.88)};

 \addplot[style={bblue,fill=bblue,mark=none}]
            coordinates {($8$, 103) ($16$, 90) ($32$, 75) ($64$, 63.25) ($128$, 47.79) ($256$, 36) ($512$, 24.78)};

        \legend{{\bf RA},  SFA}
    \end{axis}
\end{tikzpicture}
\caption{Comparison of per-job flow time for $\textsf{arr}=5$ with changing $K$ with $\sfs$ uniformly randomly among $[1 \ 2\ 4 \ \dots \ 2^{\log K}]$.}
\label{fig:Kchanging} 
\end{figure}

Next, we fix the arrival rate as $\textsf{arr}=5$, and plot the per-job flow time of {\bf RA} and SFA in Fig. \ref{fig:Kchanging} as a function of $K$, 
to demonstrate the relative performance as $K$ increases. From Fig. \ref{fig:Kchanging}, we see that the ratio of per-job flow time of SFA and {\bf RA}
increases as $K$ increases to reflect the increasing limitation of SFA compared to {\bf RA} with increasing $K$. 

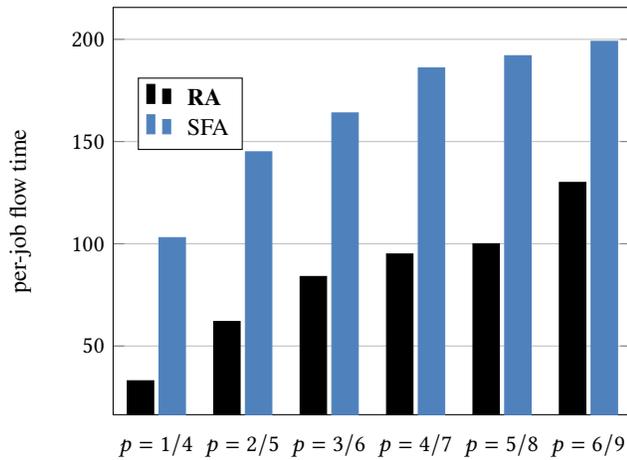
\begin{figure}
\centering
\begin{tikzpicture}
    \begin{axis}[
        width  = \columnwidth,
        height = 7cm,
        major x tick style = transparent,
        ybar,
        bar width=10pt,
        ymajorgrids = true,
        ylabel = {per-job flow time},
        symbolic x coords={$p=1/4$, $p=2/5$, $p=3/6$, $p=4/7$, $p=5/8$, $p=6/9$},
        xtick = data,
        scaled y ticks = false,
        legend cell align=left,
        legend style={
                at={(.25,.65)},
                anchor=south east,
                column sep=1ex}
    ]

     \addplot[style={black,fill=black,mark=none}]
            coordinates {($p=1/4$, 33) ($p=2/5$, 62) ($p=3/6$, 84) ($p=4/7$, 95) ($p=5/8$, 100) ($p=6/9$, 130) };
            
             \addplot[style={bblue,fill=bblue,mark=none}]
            coordinates {($p=1/4$, 103) ($p=2/5$, 145) ($p=3/6$, 164) ($p=4/7$, 186) ($p=5/8$, 192) ($p=6/9$, 199) };
            
        \legend{{\bf RA},  SFA}
    \end{axis}
\end{tikzpicture}
\caption{Comparison of per-job flow time for $K=8$, $\textsf{arr}=5$ with changing $p$, the probability $p$ of choosing $\sfs=8$.}
\label{fig:8p} 
\end{figure}

To demonstrate the dependence of results on the distribution of server need $\sfs$,  next, 
we fix $K=8$ and $\textsf{arr}=5$, and choose $\sfs=8$ for any job with probability 
$p$ and choose $s =\{ 1,2,4\}$ with equal probability of $(1-p)/3$, and plot the per-job flow time in Fig. \ref{fig:8p}. 
As $p$ increases the relative performance of SFA with respect to {\bf RA} improves, since there is less variability in job sizes with larger $p$.

\begin{figure}
\centering
\begin{tikzpicture}
    \begin{axis}[
        width  = \columnwidth,
        height = 7cm,
        major x tick style = transparent,
        ybar,
        bar width=10pt,
        ymajorgrids = true,
        ylabel = {per-job flow time},
        symbolic x coords={$K=32$, $K=64$, $K=128$, $K=256$},
        xtick = data,
        scaled y ticks = false,
        legend cell align=left,
        legend style={
                at={(.75,.65)},
                anchor=south east,
                column sep=1ex}
    ]

     \addplot[style={black,fill=black,mark=none}]
            coordinates {($K=32$, 8.44) ($K=64$, 11.8) ($K=128$, 23) ($K=256$, 53)  };
            
             \addplot[style={bblue,fill=blue,mark=none}]
              coordinates {($K=32$, 18.21) ($K=64$, 30.3) ($K=128$, 69.01) ($K=256$, 127.7)  };

 \addplot[style={rred,fill=red,mark=none}]
              coordinates {($K=32$, 2) ($K=64$, 1.5) ($K=128$, 1.2) ($K=256$, 1.12)  };

        \legend{{\bf RA},  SFA, $\cB$}
    \end{axis}
\end{tikzpicture}
\caption{Comparison of per-job flow time with $\textsf{arr}=5$ as a function of $K$ with changing $p=1/K$.}
\label{fig:rand} 
\end{figure}
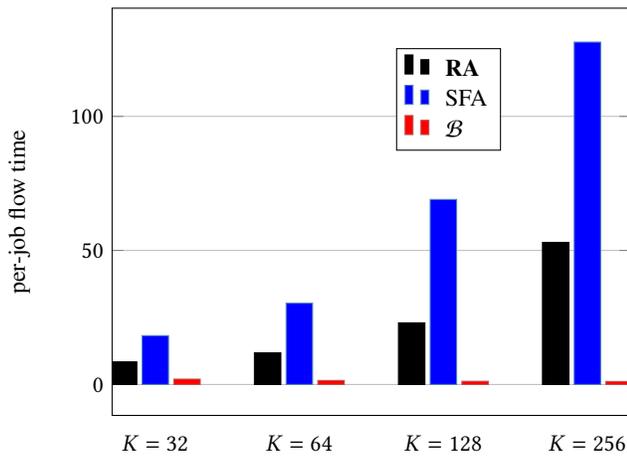

Finally in Fig. \ref{fig:rand}, we consider the input considered to derive the lower bound for all randomized algorithms; in each slot 
$K/2$ jobs with $\sfs=1$ and one job with $\sfs=K$ arrives with probability $p$ and only one job with $\sfs=K$ arrives with probability $1-p$, and 
plot the per-job flow time. We compare the performance of  {\bf RA} and SFA, and an algorithm $\cB$ that processes the $K/2$ jobs in the same slot 
as they arrive. Clearly, $\cB$ is better than   {\bf RA} and SFA for this input, and is clearly reflected in the orderwise better performance of  
$\cB$.

%
%
%
%
%

\section{Conclusions}
In this paper, we considered an important scheduling problem (multi-server jobs) for data centers where each job needs multiple servers for it to be processed, and the server demands of each job are heterogenous. Almost all prior work on this multi-server jobs problem was known for a stochastic setting, where job arrival process is assumed to be Poisson, and server needs for each job are i.i.d. In this paper, however, we considered an arbitrary input model to suit real-world situations, where job arrival times, number of jobs arriving in any slot, and the server needs of each job are arbitrary and can even be generated by an adversary. Under this very general model, when all job sizes are equal, we proposed a simple online algorithm and showed that its competitive ratio is order-wise optimal, and scales linearly in the number of total servers. One can argue that this is in fact a negative result and the power of online algorithms is rather limited compared to the offline optimal algorithm for the multi-server jobs problem. 
Hence we also considered the natural empowerment of the online setting by allowing an online algorithm to have access to more servers than the 
optimal offline algorithm. Under this resource augmentation regime, we showed an important result that a simple algorithm with access to double the number of servers than the optimal offline algorithm is as good as the optimal offline algorithm, leading to critical system design directions.

\bibliographystyle{elsarticle-num} 
\bibliography{refs}
\section{Appendix:Proof of Theorem \ref{thm:lbrand}}
\begin{proof}
We will use Yao's recipe \cite{bookvaze} to lower bound the competitive ratio of any randomized algorithm which states the following. 
For any distribution $D$ of input $\sigma$, the competitive ratio $\mu_\cR$ of any randomized algorithm $\cR$  
 is lower bounded by 
\begin{equation}\label{eq:yao}
\mu_\cR \ge \frac{\bbE_{D}\{F_{\cA^\star}(\sigma)\}}{\bbE_{D}\{F_{\opt}(\sigma)\}},
\end{equation}
where $\cA^\star$ is the optimal deterministic online algorithm for input $\sigma$ with distribution $D$.

To use this recipe, we will prescribe a distribution and then bound the performance of the optimal deterministic online algorithm and the $\opt$.

Input distribution: Let at any slot $t=1, \dots, T$, either $K/2$ jobs with $\sfs_j=1$ and $1$ job with $\sfs_j=K$ arrive with probability $p$, or $1$ job with $\sfs_j=K$ arrives with probability $1-p$.
We will choose $p=1/K$. The input after slot $T$ will depend on the number of slots used by $\cA$ where it only processes $K/2$ jobs with $\sfs_j=1$.

With this input distribution, any online algorithm $\cA$ once it gets  $K/2$ jobs with $\sfs_j=1$ in any slot will wait for $\theta=0, \dots, T$ more slots for a new set of $K/2$ jobs with $\sfs_j=1$ to arrive so that it can combine them and process them together. Choosing a small value of $\theta$ will keep the flow time of $K/2$ jobs with $\sfs_j=1$ small, while larger values of $\theta$ will reduce the wastage of server capacity (idle any server) which can be used to process more jobs with $\sfs_j=K$. 
Since we are looking for order-wise results it is sufficient to consider $\theta=O(1)$, $\theta=o(K)$ and $\theta=\Omega(K)$. The intermediate choice $\theta=o(K)$ is dominated by $\theta=\Omega(K)$ and $\theta=O(1)$ since choosing $\theta=o(K)$, the upside is the ability to combine $K$ jobs with $\sfs_j=1$ and process them together and reduce  server capacity wastage compared to $\theta=O(1)$. However, the choice of $\theta=o(K)$ only ends up increasing the flow time of jobs with $\sfs_j=1$, since the enabling event for $\theta=o(K)$, 
that the difference in consecutive slot indices where $K/2$ jobs with $\sfs_j=1$ arrive is $o(K)$, has a very small probability since $p=1/K$.  To simplify the proof we consider 
$\theta=0$ to represent $\theta=O(1)$ case, and $\theta=T$ to represent $\theta=\Omega(K)$.

We next consider  both  $\theta=0$ and $\theta=T$ (i.e. always wait until the next time $K/2$ jobs with $\sfs_j=1$ arrive) for $\cA$. 

With $\theta=0$, $\cA$ processes the $K/2$ jobs with $\sfs_j=1$ in the same slot as they arrive, and with $p=1/K$, the expected number of slots where  $K/2$ jobs with $\sfs_j=1$ arrive  in $[1:T]$ is $(T/K)$. 
As a result,  $\cA$ wastes half the server capacity for $(T/K)$ slots (in expectation), and hence the expected number of jobs with $\sfs_j=K$ remaining with $\cA$ at time $t$ is $(T/K)$. In comparison, consider an offline algorithm $\cB$ that always processes a job with $\sfs_j=1$ as soon as there are $K$ such jobs. Hence, with $\cB$, the number of jobs with $\sfs_j=K$ remaining with $\cA$ at time $T$ is $(T/2K)$. 
{\bf Input after time $T$:} No jobs arrive for interval $T+1$ to $T+(T/2K)$. Thus, at time $T+(T/2K)$, $\cB$ has no remaining jobs while $\cA$ has $(T/2K)$ remaining jobs (in expectation) with $\sfs_j=K$.
{\bf Input after time $T+(T/2K)$:} $2$ jobs with $\sfs_j=K/2$ arrive for time slots $T+(T/2K)+1, \dots, T+(T/2K)+L$. Its best for $\cA$ to process the $2$ jobs with $\sfs_j=K/2$ before processing any of its remaining jobs with $\sfs_j=K$ in terms of minimizing its flow time. Thus, the expected flow time of $\cA$ is at least $L\cdot (T/2K)$ (counting only the flow time of $(T/2K)$ remaining jobs in expectation) at time $T+(T/2K)$.

$\cB$ on the other hand has an expected flow time of at most $O(KT)$ over the period $[1:T]$, $O((T/K)^2)$ over the period $[T+1, T+(T/2K)]$ and $O(L)$ over the period  $[T+(T/2K)+1,  T+(T/2K)+L]$. Thus, the expected flow time of $\cB$ is $O(\max\{KT, (T/K)^2,L\}$.  Since $\opt$ is as good as $\cB$,
 the competitive ratio of $\cA$ with $\theta=0$ is 
 \begin{equation}\label{eq:lbrand0} \frac{\Omega(L\cdot T/(2K))}{O(\max\{KT, (T/K)^2,L\})} = \Omega(K)
 \end{equation}
  choosing 
 $L=TK$.
 
 Next, we consider $\theta=T$, i.e. $\cA$ waits to process any $K/2$ jobs with $\sfs_j=1$ until another set of $K/2$ jobs with $\sfs_j=1$ arrive next. Given that $p=1/K$, the expected wait time to get two slots where $K/2$ jobs with $\sfs_j=1$ arrives is $K$. Thus, over the time horizon of $T$, the expected flow time of $\cA$ is $\Omega(KT)$, since the expected flow time of $K/2$ jobs with $\sfs_j=1$ that wait for a new set of $K/2$ jobs with $\sfs_j=1$ to arrive is $\Omega(K^2)$, and there are $\Omega(T/K)$ slots in which $K/2$ jobs with $\sfs_j=1$ arrive.
 In comparison,  consider an offline algorithm $\cB$ that processes the $K/2$ jobs with $\sfs_j=1$ in the same slot as they arrive. This way, the expected flow time of $\cB$ for all the job with $\sfs_j=1$ in interval $[1:T]$ is $\Omega(T)$. But with $\cB$, the expected number of jobs with $\sfs_j=K$ remaining at time $T$ is $T/K$. The expected flow time of $\cB$ of jobs with $\sfs_j=K$ that are processed  in $[1:T]$ is $O(T)$ and the ones processed after $T$ is $O( T^2/K+ (T/K)^2)$. Thus, the overall expected flow time of $\cB$ is $O(\max \{T, (T^2/K)\})$. Since $\opt$ is as good as $\cB$, the competitive ratio of $\cA$ with $\theta=T$ is 
 \begin{equation}\label{eq:lbrandT}\frac{\Omega(KT)}{O(\max \{T, (T^2/K)\})} = \Omega(K),
 \end{equation} 
 choosing $T=\Theta(K)$.
 
Thus, from \eqref{eq:lbrand0} and \eqref{eq:lbrandT}, for any value of $\theta$, the competitive ratio of $\cA$ is $\Omega(K)$, and hence for $\cA^\star$ the competitive ratio  is $\Omega(K)$. Thus, from \eqref{eq:yao}, we get the result.

\end{proof}
\section{Proof for Theorem \ref{thm:ubunequalsize}}
Let the algorithm {\bf RA-Size} be denoted as $\cA$. Job $j$ is defined to belong to class $a$ at time slot $t$ if its effective size $w'_j(t) = w_j(t) \cdot \sfs_j \in [2^a, 2^{a+1}]$ 
for $a=0, 1, \dots, \log (Kw_{\max})-1$. 

Recall the definition of $t^-$ and $t^+$ from Definition \ref{defn:tminus}.
For $\cA$, let $R(t^-)$ be the set of outstanding/remaining jobs at slot $t^-$ with $n(t) = |R(t^-)|$, and $n_a(t)$ is the number of remaining jobs with $\cA$ belonging to class $a$ at $t^-$. Moreover, for $\cA$ let $W(t) = \sum_{j \in R(t^-)} w_j(t) \sfs_j$ be the volume (the total outstanding workload) at slot $t^-$. 
 
 Consider the potential function \begin{equation}\Delta W(t) = W(t) - W^\opt(t),\end{equation}  that represents the difference in volume between $\cA$ and the  $\opt$.


 For any quantity denoted by $Q\in \{W, \Delta W\}$, $Q_{\ge \ell, \le h}$ means the respective quantity when restricted to jobs of classes 
  between $\ell$ and $h$, and $Q_{x} = Q_{\ge x, \le x}$.
  
  \begin{definition} The system is defined to be {\bf full} at slot $t$ if all the $K$ servers are occupied by $\cA$. The set of slots when the system is full is denoted as $\cT_f$. 
  If the system is not full at slot $t$, then it is defined to be {\bf relaxed}, and the set of slots when the system is relaxed is denoted as $\cT_r$.
  \end{definition}
  
 \begin{lemma}\label{lem:relaxunequalsize}
  If the system is relaxed at slot $t$, i.e. if $t\in T_r$, then $n(t)\le K$.
\end{lemma} 
Proof is identical to Lemma \ref{lem:relax}.


To complement Lemma \ref{lem:relaxunequalsize}, we have the following lemma for bounding the number of outstanding jobs with $\cA$ at slot $t \in \cT_f$ belonging to classes between $a_1$ and $a_2$.
\begin{lemma}\label{lem:fullunequalsize}
   For $t\in \cT_f$ 
   $$n_{\ge a_1, \le a_2}(t) \le (a_2-a_1+2)(K-1) + 2 n^\opt_{\le a_2}(t)$$
  \end{lemma}
  Since there are at most $\log (Kw_{\max})$ classes, summing over all possible $a_1,a_2$, we have 
  \begin{equation}\label{eq:finaloutstandingjobsunequalsize}
n(t) \le (K-1)\log (Kw_{\max})+ 2 n^\opt(t).
\end{equation}
 
 Next, using Lemma \ref{lem:relaxunequalsize} and \eqref{eq:finaloutstandingjobsunequalsize} and the following simple observations $F_\cA  = \sum n(t) $, and $|\cT_f| + |T_r| \le \sum_{j \in \cJ} w_j$,  $F_\opt \ge \sum_{j \in \cJ} w_j$, we complete the proof of Theorem \ref{thm:ub}. Proof of Lemma \ref{lem:fullunequalsize} is provided thereafter.

\begin{proof}[ Proof of Theorem \ref{thm:ubunequalsize}]
  \begin{align*}
  F_\cA& \stackrel{(a)}= \sum_t n(t) ,   \\
  &  \stackrel{(b)}= \sum_{t\notin \cT_f} n(t)  +  \sum_{t\in \cT_f} n(t) ,   \\
&  \stackrel{(c)}\le \sum_{t\notin \cT_f} (K-1) +  \sum_{t\in \cT_f}  (K-1)\log (Kw_{\max})+ 2 n^\opt(t),   \\
&  \le (|\cT_f| + |T_r|)(K-1)\log (Kw_{\max}) + 2 \sum_{t} n^\opt(t) , \\
&  \stackrel{(d)}\le  (K-1)\log (Kw_{\max})  \sum_{j \in \cJ} w_j+ 2 \sum_{t} n^\opt(t) , \\
&  \stackrel{(e)}\le (K-1)\log (Kw_{\max}) F_\opt + 2 \sum_{t} n^\opt(t) , \\
&  =  (K+1)\log (Kw_{\max})F_\opt
\end{align*}
where $(a)$ follows from the definition of flow time, and $(b)$ follows by partitioning time into sets $\cT_f$ and $T_r$. 
Lemma \ref{lem:relaxunequalsize} and \eqref{eq:finaloutstandingjobsunequalsize} together imply $(c)$, while $(d)$ follows since  $(T_r + \cT_f) \le \sum_{j \in \cJ} w_j$, and because trivially $F_\opt \ge  \sum_{j \in \cJ} w_j$ we get $(e)$.
\end{proof}

Next, we work towards proving Lemma \ref{lem:full}.
\begin{definition}
For some $t\in \cT_f$, let
${\hat t} < t$, be the earliest slot such that $[{\hat t}, t) \in \cT_f$, i.e. for all slots $[{\hat t}, t)$ all servers are busy with $\cA$.
  During interval $[{\hat t}, t)$, the latest slot at which a job belonging to class greater than $a$ is processed is defined as $t_a$. 
  We let $t_a = {\hat t}-1$, if no job with class greater than $a$ is processed in $[{\hat t}, t)$.
  \end{definition}

  With these definitions, we have the following intermediate result.
  \begin{lemma}\label{lem:driftrelation}
  For $t\in \cT_f$, $\Delta W_{\le a}(t) \le  \Delta W_{\le a}(t_a+1)$.
\end{lemma}
Proof is identical to Lemma \ref{chap:mssched:lem:deltav1}, since all servers are busy throughout the interval $[t_a+1, t)$ with $\cA$ processing 
jobs with class at most $a$ and hence $\cA$ reduces the volume $W_{\le a}$ by maximal 
amount of $K$ at any slot in  $[t_a+1, t)$.

Next Lemma is the first place where jobs having unequal sizes matters.

\begin{lemma}\label{lem:driftrelation2}
For $t\in \cT_f$, $\Delta W_{\le a}(t_a+1) \le (K-1) 2^{a+1}$.
\end{lemma}
\begin{remark}
The analogous result when job sizes are equal, (Lemma \ref{chap:mssched:lem:deltav2}), is significantly better.
\end{remark}

\begin{proof}

Case I: $t_a={\hat t}-1$. Thus, no job with class more than $a$ is processed by $\cA$ in $[{\hat t}, t)$. Since ${\hat t}-1\in \cT_r$, we get that the total number of jobs with $\cA$ at both the start and end of slot 
${\hat t}-1$ with class at most $a$ is at most $K-1$. Thus, $W_{\le a}(t_a^+) = (K-1) 2^a$. This is true since otherwise $\cA$ would have processed some subset of jobs with class at most $a$ while completely fitting the $K$ servers.
Moreover, the set of newly  arriving jobs in slot $t_a+1$ is identical for both $\cA$ and the $\opt$. Thus, we get $\Delta W_{\le a}(t_a+1) \le W_{\le a}(t_a^+)\le K-1$.


Case II: $t_a>{\hat t}$. If $\cA$ is processing a job of class more than $a$ at slot $t_a$ this means that the total number of jobs at slot $t_a^-$ with class at most $a$ is at most $K-1$.This is true since otherwise a subset of the $K$ or more jobs $\cA$ has of class at most $a$ would exactly fit the $K$ servers, and the window set $S_{i^\star}$ 
chosen for processing at slot $t$, will consists entirely of jobs belonging to class  at most $a$.  
Since the effective size of any job belonging to class $a$ is at most $2^{a+1}$, therefore, we get that 
  $W_{\le a}(t_a^+)\le (K-1) 2^{a+1}.$
  Moreover, the set of newly  arriving jobs in slot $t_a+1$ is identical for both $\cA$ and the $\opt$, thus, we get $\Delta W_{\le a}(t_a+1) \le W_{\le a}(t_a^+)\le (K-1)2^{a+1}$.

\end{proof}

Combining Lemma \ref{chap:mssched:lem:deltav1} and \ref{chap:mssched:lem:deltav2}, we get the following result.

\begin{lemma}\label{lem:drift}
  For $t\in \cT_f$, $\Delta W_{\le a}(t) \le  K-1$.
\end{lemma}

Using Lemma \ref{lem:drift}, we complete the proof of Lemma \ref{lem:full}, that connects the number of remaining jobs with the algorithm and the $\opt$, belonging to a certain set of classes.

\begin{proof}[Proof of Lemma \ref{lem:full}]
\begin{align}\nn
   \sum_{a=a_1}^{a_2} n_a(t)& \stackrel{(a)}\le \sum_{a=a_1}^{a_2}\frac{W_a(t)}{2^a},  \\\nn
  &\stackrel{(b)} =  \sum_{a=a_1}^{a_2}\frac{\Delta W_a(t) +  W^\opt_a(t)}{2^a}, 
    \end{align}
  \begin{align}\nn
  &=  \sum_{a=a_1}^{a_2}\frac{\Delta W_{\le a}(t)  - \Delta W_{\le a-1}(t)}{2^a} + \sum_{a=a_1}^{a_2}\frac{ W^\opt_a(t)}{2^a},  \\\nn
    & \stackrel{(c)}\le    \frac{\Delta W_{\le a_2}(t)}{2^{a_2}} + \sum_{a=a_1}^{a_2-1}\left(\frac{\Delta W_{\le a}(t) }{2^{a}}- \frac{\Delta W_{\le a}(t) }{2^{a+1}}\right) - 
  \frac{\Delta W_{\le a_1-1}(t)}{2^{a_1}} \\ \nn
 & \quad \quad  + \sum_{a=a_1}^{a_2}\frac{ W^\opt_a(t)}{2^a}, \\ \nn
  &  \stackrel{(d)}\le    \frac{\Delta W_{\le a_2}(t)}{2^{a_2}} + \sum_{a=a_1}^{a_2-1}\frac{\Delta W_{\le a}(t) }{2^{a+1}} - 
  \frac{\Delta W_{\le a_1-1}(t)}{2^{a_1}} + 2 n^\opt_{\ge a_1, \le a_2}(t), \\ \nn
     &  \stackrel{(e)}\le 2(K-1) +  \sum_{a=a_1}^{a_2-1}(K-1) + 
  \frac{ W^\opt_{\le a_1-1}(t)}{2^{a_1}} + 2 n^\opt_{\ge a_1, \le a_2}(t), \\\nn
   &  \stackrel{(f)}\le (a_2-a_1+2)(K-1)  + 2n^\opt_{ \le a_1-1}(t) + 2 n^\opt_{\ge a_1, \le a_2}(t), \\\nn
   &  \le  (a_2-a_1+2)(K-1) + 2 n^\opt_{\le a_2}(t)
\end{align}
where $(a)$ follows from the definition of $W_a(t)$ as the total remaining volume of jobs  and a job belonging to class $a$ at slot $t$ has effective size $w' \in [2^a, 2^{a+1}]$, while $(b)$ follows from the definition of $\Delta W_a = W_a - W^\opt_a$. To get $(c)$ we separate the telescopic sum over $a_1$ to $a_2$ into three parts, 
$a_2$, $a_1$ to $a_2-1$ and $a_1-1$. On the second part we use the fact that $\Delta W_{\le a}(t)\le 2 \Delta W_{\le a-1}(t)$. Inequality $(d)$ follows from the  follows from the definition of $W_a(t)$ as the total remaining volume of jobs  and a job belonging to class $a$ at slot $t$ has effective size $w' \in [2^a, 2^{a+1}]$. Inequality $(e)$ follows by applying Lemma \ref{lem:driftrelation2} on the first two terms separately, and for the third term use the property that  $- \Delta W_{\le a_1-1}(t) \le  W^\opt_{\le a_1-1}(t)$. Using the fact that volume of a job of class $a$ is at most $2^{a+1}$  on the third term of $(e)$, we get inequality $(f)$.

\end{proof}

\end{document}